\documentclass[preprint,5p,12pt,twocolumn]{elsarticle}
\usepackage{lineno}
\journal{[...]}
\usepackage{amssymb}

\usepackage[acronym,toc,nonumberlist,style=super,nogroupskip]{glossaries}
\newacronym{ADCS}{ADCS}{Attitude Determination and Control System}
\newacronym{CPU}{CPU}{Central Processing Unit}
\newacronym{ESA}{ESA}{European Space Agency}
\newacronym{GNU}{GNU}{GNU’s Not UNIX}
\newacronym{GSL}{GSL}{GNU Scientific Library}
\newacronym{JAXA}{JAXA}{Japan Aerospace Exploration Agency}
\newacronym{JPL}{JPL}{Jet Propulsion Laboratory}
\newacronym{NASA}{NASA}{National Aeronautics and Space Administration}
\newacronym{OD}{OD}{Orbit Determination}
\newacronym{PSF}{PSF}{Point Spread Function}
\newacronym{RAM}{RAM}{Random Access Memory}
\newacronym{rms}{RMS}{\emph{Root Mean Square}, $1\,\sigma$}
\newacronym{MCC}{MCC}{Multiple Cross-Correlation}
\newacronym{UKF}{UKF}{Unscented Kalman Filter}
\newacronym{TRL}{TRL}{Technology Readiness Level}

\usepackage{xcolor, soul}
\sethlcolor{yellow}
\newcommand{\abbreviations}[1]{%
  \nonumnote{\textit{Abbreviations:\enspace}#1}}

\newcommand\mcc{\gls{MCC}}
\newcommand\ukf{\gls{UKF}}
\newcommand\ADCS{\gls{ADCS}}
\newcommand\TRL{\gls{TRL}}

\begin{document}

\begin{frontmatter}
\abbreviations{\\
ADCS, Attitude Determination and Control System;\\
CPU, Central Processing Unit;\\
ESA, European Space Agency;\\
GNU, GNU’s Not UNIX;\\
GSL, GNU Scientific Library;\\
JAXA, Japan Aerospace Exploration Agency;\\
JPL, Jet Propulsion Laboratory;\\
MCC, Multiple Cross-Correlation;\\
NASA, National Aeronautics and Space Administration;\\
OD, Orbit Determination;\\
PSF, Point Spread Function;\\
RAM, Random Access Memory;\\
RMS, Root Mean Square ($1\,\sigma$);\\
TRL, Technology Readiness Level;\\
UKF, Unscented Kalman Filter.
}



\title{Autonomous Orbit Determination for a CubeSat Cruising in Deep Space}

\author[lesia_psl]{Boris Segret\corref{cor1}}
\ead{boris.segret@observatoiredeparis.psl.eu}
\author[lesia_psl,sorbonne,pdiderot]{Benoît Mosser}
\address[lesia_psl]{LESIA, Observatoire de Paris - PSL, 5 pl. Jules Janssen, 92195 Meudon Cedex, France}
\address[sorbonne]{Sorbonne Université, 21 rue de l'école de médecine, 75006 Paris, France}
\address[pdiderot]{Université de Paris, 85 boulevard Saint-Germain 75006 Paris, France}
\cortext[cor1]{Corresponding author.}

\begin{abstract}
CubeSats have become a meaningful option for deep-space exploration, but their autonomy must be increased to maximize the science return while limiting the complexity in operations. We present here a solution for an autonomous orbit determination in the context of a CubeSat cruising in deep space. The study case is a journey from Earth to Mars. An optical sensor at CubeSat standard is considered. The image processing is added to extract the direction of distant celestial bodies with 0.2\," accuracy: it consists of a \mcc\ algorithm that uses bright stars in the background of the images. Then, an \ukf\ is built to perform an asynchronous triangulation from the successive directions of the celestial bodies. The \ukf\ meets the expected performance in contexts where linear approximations are not possible. The orbit reconstruction reaches a $3\,\sigma$ accuracy of 30\,km in the middle of the Earth-Mars cruise. Additionally, the \gls{CPU} cost of the filter is assessed at less than 1 second per iteration with a typical CubeSat hardware. It is ready for further improvements in terms of new observables associated with data fusion, quicker convergence and attitude control savings.
\end{abstract}

%


\end{frontmatter}


\section{CubeSats for deep-space need more autonomy}\label{science}

Satellites of a few kilograms have become very popular since the success of the form factor called CubeSat, a satellite that meets the CubeSat Design Specification \cite{CalPoly2009}. The disruption permitted in Low Earth Orbit is now addressing new scientific needs for space exploration. Indeed, the CubeSats may be a good option for exploration missions, provided their scientific added-value will be much higher than their cost at building, integrating to a host mission, and operating. A key approach is to develop their autonomy, in particular for navigation. The few experiments that were carried out in space are addressed in section \ref{past}, with a selection of mission concepts and our previous results. Then we present in section \ref{scCase} our science case that has been the motivation for the development of the presented solution.

After this introduction, we specify our whole algorithm for an optical orbit determination on board a nanosatellite in cruise, without assistance from the ground. In section \ref{ot}, an optical sensor compatible with the CubeSat standard is described and an image processing called \mcc\ is specified. Section \ref{at} defines the elements of a \ukf, its models and its updating process. The results for our study case in cruise context are presented in section \ref{results}, in terms of accuracy, sensitivity and CPU cost. Then, the discussion in section \ref{next} addresses the potential for further improvements and adaptations to data fusion and proximity operations.

\subsection{Past and current attempts}\label{past} 

Satellites called CubeSats have been adopted for 20 years in multiple kinds of space missions in Low Earth Orbit. Japan had popularized an ultra-small satellite format already in the 90s, called CanSat, for suborbital experiments (e.g. \cite{Matunaga2000}, \cite{Sako2001}). Then, California Polytechnic State University, San Luis Obispo and Stanford University's Space Systems Development Lab created the CubeSat standard in 1999 that greatly stimulated the emergence of the economic \emph{New Space} sector. Such satellites successfully addressed multiple scientific missions or in-orbit demonstrations, stimulating the development of reliable subsystems at this scale, including automatic maneuvers, e.g. in Guo 2019 \cite{Guo2019}. Autonomous navigation, that includes both orbit determination and correction maneuvers, is one of the key technologies expected for this form factor, targeting in particular the management of swarms and constellations, e.g. in Yoshimura 2018 \cite{Yoshimura2018}.

For Deep-space missions, CubeSats become a meaningful option, especially for proximity operations as auxiliary probes of a larger mothercraft. Deep-space missions have already included auxiliary probes, like ESA's mission \textsc{Rosetta} with \textsc{Phil\ae}, or JAXA's mission \textsc{Hayabusa 2} with German-French \textsc{Mascot} along with many other tiny probes. The launch of the twin CubeSats \textsc{MarCO} by NASA's JPL in 2018 from the mothercraft carrying \textsc{InSight} was a cornerstone: after jettisoning at the beginning of the journey to Mars, they reached the red planet successfully, which granted the CubeSat standard a proof of deep-space sustainability.

However, none of these probes ever navigated autonomously, raising the question of their added-value compared to their cost in terms of telecommunications and operations from the ground.

There was not any actual autonomous navigation solution for deep-space missions. The fact seems to result from a tradition that keeps a quasi-continuous link between ground and spacecraft, allowing the orbit determination and any trajectory correction maneuvers to be computed on the ground and transmitted to the spacecraft. Although several attempts were performed, publicly available documentation is sparse. \textsc{Deep Space 1} by NASA had an experiment for autonomous navigation: Riedel et al. 2000 \cite{Riedel2000} presented a fair accuracy down to 150\,km and 0.2\,m/s ($1\,\sigma$) after processing images for 25 days of the cruise. However, the accuracy was worse sometimes, and the solution remained a demonstration project only, without any operational use. NASA's mission \textsc{Deep Impact} embedded two ``AutoNav" solutions \citep{Kubitschek2006}: the impactor was released 24 hours before the impact and ran an auto-guidance algorithm based on local short-term physical laws, applying lateral $\Delta V$ to keep the target in sight and to hit it. The mothercraft ran a so-called ``AutoNav" consisting of controlling the line of sight and the functional modes of its instruments to the target during the event, hence nothing related to the navigation per se. Some studies have suggested the use of pulsar-based triangulation, with either radio pulsars or X-ray pulsars. Solutions are not currently feasible for nanosatellites: Martindale et al. 2015 \cite{Martindale2015} reminded that the sensors for X-ray pulsars are large and must be pointed in three different directions with an on-board time-stamping at $10^{-12}$\,s accuracy, with regular updates of the sources catalog (provided from the Earth). For radio pulsars, Jessner 2015 \cite{Jessner2015} showed that the antennas and their pointing are beyond nanosatellite capacities, among other constraints, for an expected accuracy of 60 to 180\,km at best.

The \textsc{Deep Space 1} experiment embedded an image processing called ``\mcc" for Multiple Cross-Correlation. It is a center-finding algorithm that exploits the fact that all objects undergo the same movement during exposure. Their images are then identically smeared as the same pattern. The technique gets rid of non-stabilized pointing and allows even faint objects to be located relative to one another with a claimed accuracy as low as 0.1 pixels (Riedel et al. \cite{Riedel1997}, Vaughan et al. \cite{VAUGHAN2013}). Additionally, we note that the distances on the image from one given object to all others are also correlated, allowing an enhancement of the \mcc\ that we specify here in full details.

While the interest for autonomous navigation is debatable for traditional spacecrafts, it is obviously high for deep-space CubeSats in order to save cost and complexity in operations. The twin CubeSats \textsc{MarCO} were entirely navigated from the ground in the same way as any other deep-space mission. Tortora and Di Tana 2019 \cite{Tortora2019} claimed an autonomous fly-by with on-board orbit determination for the Italian CubeSat \textsc{LiciaCube} that shall accompany NASA's DART spacecraft to impact Dimorphos in 2022. The solution referred to a study by Modenini et al. 2018 \cite{Modenini2018} for in-situ navigation that is based on images of an ellipsoid-shape asteroid. Then the study should be adapted to a fast fly-by for \textsc{LiciaCube}. No results could be found at the time of writing. An Italian precursor \textsc{Lumio} was presented by Franzese et al. 2018 \cite{Franzese2018} with a full Moon imaging technique to maintain a 14-day periodic halo orbit at Earth-Moon L2 Lagrangian point (distance to the Moon is then $\sim35\,000$\,km). Simulations showed that the mean error in orbit determination was lower than 10\,km. The technique belongs to the horizon-based navigation family that is promising in contexts of proximity operations where resolved images are available for known celestial objects. On-board orbit determination for ESA's technology demonstrator \textsc{M-Argo} was announced by Walker et al. 2015 \cite{Walker2015} but not published yet. Whether an on-board or ground orbit determination would be designed for ESA \textsc{Hera}'s auxiliary CubeSats at Didymos and Dimorphos in 2027 has been unclear.

These numerous examples, although not exhaustive, show the timeliness of the study of on-board orbit determination for deep space nanosatellites. Our approach is an asynchronous triangulation of optical directions of foreground bodies. The initial assessments were presented in 2017\,\cite{Segret2017}. In 2018\,\cite{Segret2018}, we provided an initial formulation of an enhanced \mcc\ and reached an orbit reconstruction with $\sigma \sim 100$\,km in cruise with a batch, linear Kalman Filter and Monte-Carlo simulations, assuming that an optical accuracy could be improved by the enhanced \mcc\ at 0.2\,$''$. In 2019\,\cite{Segret2019b}, the Monte-Carlo simulations were replaced by a covariance analysis to study the sensitivity of the method. Here, we reformulate the enhanced \mcc\ and justify its performance, then we recapitulate the whole process, while adopting a sequential \ukf\ as a better estimator.

\subsection{Scientific needs for a study case}\label{scCase}

Multiple scientific goals are already identified that cannot be addressed with traditional deep-space missions. Interplanetary space weather is an example. When NASA's \textsc{MSL Curiosity} traveled from Earth to Mars in 2011-2012, its instrument RAD (Radiation Assessment Detector) was turned on and gathered a full set of data during the cruise, covering several solar events that occurred in the period. Posner et al. \cite{Posner2013} in 2013 confirmed a so-called Hohmann-Parker effect from the data-set and, hence, the interest to deploy a multi-site measurement system at the solar system scale that would take advantage of every single interplanetary mission to probe the solar wind.

However, RAD, and most likely any instrument, could not be oriented properly to meet specific scientific needs  during the cruise. But if a CubeSat is jettisoned from the interplanetary mission, it can be independent of it and then solve the issue. This measurement concept was presented first in 2013 as a project called \textsc{Birdy} by Vannitsen et al. \cite{Vannitsen2013a} and further documented \cite{Vannitsen2017b} in 2017. The concept in \textsc{Birdy} for autonomous navigation also inspired a new scientific application for space geodesy: in proximity operations at an explored asteroid, Hestroffer 2017 \cite{Hestroffer2017} suggested that the CubeSat be a radio-science probe, jettisoned in-situ from a mothercraft. With autonomous navigation, it could fly multiple times at low altitudes and low velocity for efficient radio-science measurement, while avoiding costly ground operations.

\section{Sensor: the Object Tracker}\label{ot}

\begin{figure}[t]\centering
   \includegraphics*[width=\linewidth]{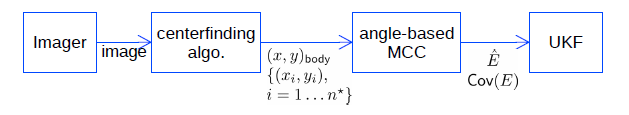}
   \caption{\label{fig:blockd}Image processing. A centerfinding algorithm extracts the locations of the body and the stars on the image, the angle-based correlation derives body's precise correction of direction.}
\end{figure}

The orbit determination we present uses on-board optical measurements. The elementary measurement consists of an image of an unresolved foreground celestial body in front of known background stars. The region of the sky where the body shall be searched for is estimated from a pre-computed trajectory of the probe that is stored on-board, before the flight, as a ``reference'' trajectory $T_R$. Then the image is processed on-board (Fig.\,\ref{fig:blockd}): a centerfinding algorithm extracts the raw locations of the searched body $(x,y)$ and of $n^\star$ expected known stars $\{(x_i,y_i),i=1\dots n^\star\}$ around it. Then, the enhanced \mcc\ derives a correction $\hat E$ of the measured direction of the body $(x,y)$, as seen from the ``actual'' trajectory $T_A$, with a variance-covariance matrix (or ``covariance" in short). Next, the corrected direction, along with its covariance, feed the Kalman filter, as presented in section \ref{at}.

\subsection{Platform sensor}\label{sensor}

The platform is assumed to be in line with nowadays CubeSat performance in Low Earth Orbit. Hence, an attitude determination and control system is anticipated with at least 0.5\,$^\circ$ in pointing accuracy and 30\," in attitude estimate accuracy (the subsequent requirements on the platform, for instance, wheel-off-loading, are beyond the scope of this paper). Then an imager is anticipated with a several-degree field of view, making sure that the platform can point and keep pointing to a direction of the sky where a celestial body is expected. The inertial pointing is expected to be maintained within 5\,$''$ (assumed to be a pixel's angular half-size) during integration times up to a second.

The approximate direction, where to look, is known from $T_R$. The celestial body is searched within a Region of Interest (RoI) around this direction by a centerfinding algorithm. The maximum parallax between $T_R$ and $T_A$ must keep the body in the RoI, hence defining the maximum acceptable shift of $T_A$ from $T_R$, i.e., the extent of the flight domain around $T_R$ for the method to work. Numerical applications are given in section \ref{results} and figure \ref{fig:E2M}.

\begin{figure}[t]\centering
   \includegraphics*[width=\linewidth]{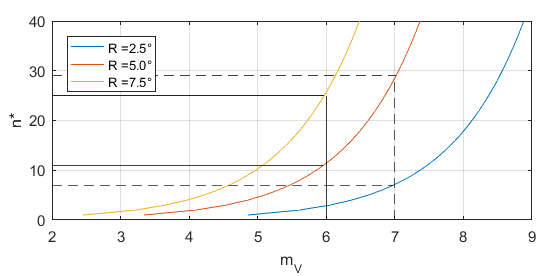}
   \caption{\label{fig:OTnEtoiles}Average number of visible stars $n^\star$ as a function of threshold magnitude $m_V$ for conic fields of view (half-angle $R$).}
\end{figure}

The imager performance must also allow images of several background stars in association with the celestial body.

The average number of stars per surface unit on the celestial sphere was computed from the basis of 6000 stars being visible up to magnitude $m_V=6$, their number being multiplied by 2.5 per magnitude. Fig.\,\ref{fig:OTnEtoiles} shows that a sensitivity between magnitudes $m_V=$ 6 and 7, as claimed by various imagers and star trackers of the \emph{New Space}, shall allow the imaging of an average number of stars from 11 to 29 within a conic field of view of 5$^\circ$ half-angle.

Then in mission preparation, a selection of celestial bodies to observe is made based on their properties as seen from $T_R$, considering their observability (defined by their magnitude and their angle to the Sun) and the accuracy of their ephemerides. Candidates are the planets if unresolved, or the planets' bright moons otherwise, or bright asteroids. Planets, if optically resolved, introduce an additional complexity, which is not studied in this paper. Ephemerides of the planets and their major moons are known with sub-kilometer accuracy. They must be interpreted with an on-board time reference. Space-grade atomic clocks of a few cm$^3$ are available\footnote{e.g. Chip Scale Atomic Clock, by Microsemi Corp., Microchip Technology Inc.} with accuracy and stability at the level of $10^{-10}$, which could limit error below 0.1\,s after one year. Also, the light travel time can be estimated on-board based on $T_R$, with an error below 0.1\,s for a shift of $T_A$ up to 30\,000\,km from $T_R$. Considering typical velocities of 30\,km/s in the solar system, there errors translate into an error of $\sim4\,$km on the estimated locations of the foreground bodies, well below the error in the measurement of optically unresolved bodies.

The asteroids are likely unresolved and can be selected as candidates for navigation if their ephemerides are well known. The release of ESA's Gaia catalog DR2 allows orbit fitting for more than 14\,099 solar system objects (\citep{GaiaCollaboration2018}, 2018). For those with magnitude in G band $10 < G \leq 13$ (as seen by \textsc{Gaia}), individual observations are always better than 0.025$''$ in accuracy on one axis and 0.005'' on the other (bright objects with $G \leq 10$ were not included in this release). From such observations, one can expect accurate ephemerides for bright asteroids if they can be imaged with a CubeSat sensor. Moreover, the next release DR3 of ESA's Gaia catalog should add observations for 350\,000 solar system objects, further enlarging the basis for well-known ephemerides. As a result, multiple asteroids, in addition to planets' moons, can likely be considered available for orbit determination.

The availability of foreground objects is a critical task of the mission preparation and an active domain of research (e.g. Broschart et al 2019 \cite{Broschart2019} or Franzese and Topputo 2020 \cite{Franzese2020}), which is not part of the present work. Here, it is assumed that a CubeSat imager can picture a selection of foreground celestial bodies with sufficient background stars at any time.

\subsection{Multiple Cross-Correlation (MCC) enhanced}\label{mcc}

\begin{figure}[t]\centering
   \includegraphics[width=0.95\linewidth]{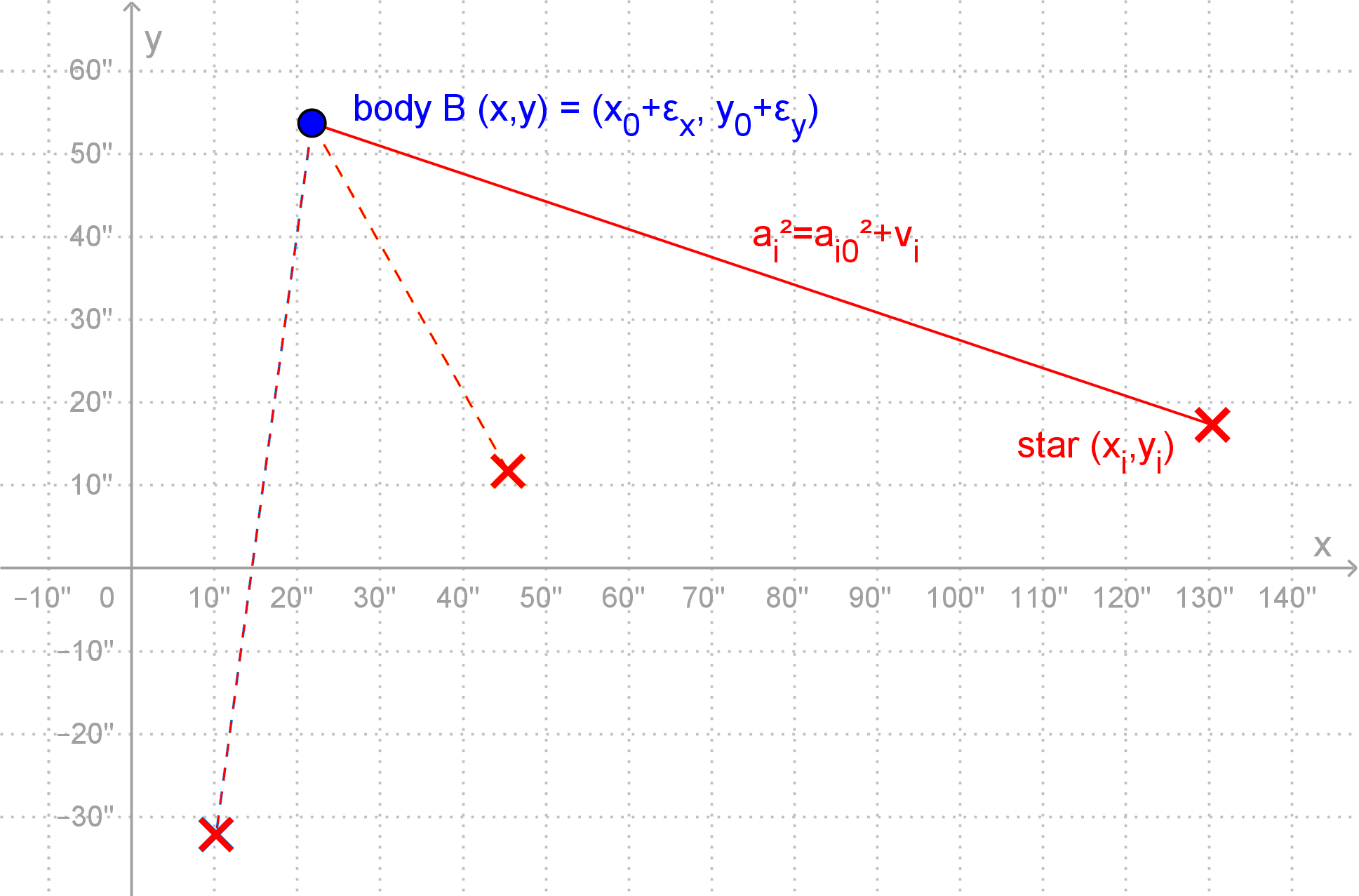}
   \caption{\label{fig:OTpfocal}Notations in the field of view for the \mcc: ``B" is the foreground body's image, ``X" are images of known background stars.}
\end{figure}

Our orbit determination uses measurements of the directions of celestial bodies improved by an angle-based correlation specified here. It is assumed that the raw directions of the objects (see Fig.\,\ref{fig:blockd}) are first extracted by a centerfinding algorithm, with an accuracy noted $\sigma_{\textsf{in}}$. For instance, the \mcc\ developed by Vaughan and Riedel \cite{Riedel1997} can produce such inputs with $\sigma_{\textsf{in}} \sim 0.1$\,pixel. These directions are then taken as ``inputs'' for our additional algorithm that correlates the multiple angular distances of objects with one another. As such, it is an enhancement of the original \mcc.

An object of interest, namely a celestial body in the foreground is located in the image with expected known stars in the background. The notations are summarized in figure \ref{fig:OTpfocal}: the true location of B, image of the celestial body, is $(x_0,y_0)$ but it is measured, with an error $(\varepsilon_x,\varepsilon_y)$, at $(x,y)=(x_0+\varepsilon_x,y_0+\varepsilon_y)$. For the $n^\star$ stars, their locations are at $(x_{0i},y_{0i})_{i=1..n^\star}$ as known from a catalog after projection on the focal plane, but they are measured in the image at $(x_i,y_i)_{i=1..n^\star}$. The square of the true angular distance between the body and a given star is:
\begin{equation}\label{eq:ang_dist}
 a_{0i}^2=(x_0-x_{0i})^2+(y_0-y_{0i})^2 .
\end{equation}

An estimate of Eq.\ref{eq:ang_dist} is given by the measurements:
\begin{equation}\label{eq:ang_meas}
 a_i^2=(x-x_i)^2+(y-y_i)^2.
\end{equation}

Hence, $a_i^2=a_{0i}^2+\nu_i$, where $\nu_i$ is assumed to be a zero-mean Gaussian random variable, built from the measurements $(x_i,y_i)$. As seen from the multiple measurements $a_i^2$ in a single image, $(\varepsilon_x,\varepsilon_y)$ is a bias. Its value corresponds to the best fit of the positions of all objects in the field of view. It can be estimated by developing Eq.\,\ref{eq:ang_dist} with $\varepsilon_x$ and $\varepsilon_y$, comparing with Eq.\,\ref{eq:ang_meas}, then building a linear system of equations, at first order:

\[\begin{array}{l}
\begin{pmatrix}\vdots \\ a_i^2-(x-x_{0i})^2-(y-y_{0i})^2 \\ \vdots\end{pmatrix} =\\
\begin{pmatrix}
\vdots & \vdots \\ -2(x-x_{0i}) & -2(y-y_{0i}) \\ \vdots & \vdots
\end{pmatrix}\begin{pmatrix}\varepsilon_x \\ \varepsilon_y\end{pmatrix}
+\begin{pmatrix}\vdots \\ \nu_i \\ \vdots\end{pmatrix}
\end{array}\]
or

\[
A = CE + N,~\textsf{where}~N \sim \mathcal{N}\left(0,W\right),
\]
where the notation $\sim \mathcal{N}(0,W)$ defines the left-hand part as a zero-mean Gaussian random variable with covariance $W$. The system can be inverted by the weighted least squares method, with a weight matrix W, that provides an estimate for $E=(\varepsilon_x,\varepsilon_y)^T$ and its covariance:

\begin{equation}
\label{eq:OTsol}\begin{aligned}
\widehat{E} &= (C^T W C)^{-1} C^T W A
\\
\textsf{Cov}(E) &= (C^T W C)^{-1}.
\end{aligned}
\end{equation}

The matrix $(C^TWC)$ has rank 2 and its elements are easy to compute provided $W$ is known. $W$ is the diagonal matrix of the \emph{inverses} of $\textsf{Var}(\nu_i)$. We assume that the relative orientation of the field of view is known with respect to absolute directions, which supposes some initial processing to identify a set of reference stars (with a minimum of two stars recognized in the field of view). We then consider that measuring the photocenter of a star image is a process dominated by a uniform, zero-mean, Gaussian error with standard deviation $\sigma_{in}$, expressing the performance of the centroiding algorithm:

\[\begin{aligned}
(x_i-x_{0i}) &\sim \mathcal{N}(0,\sigma_{in}^2)
\\
(y_i-y_{0i}) &\sim \mathcal{N}(0,\sigma_{in}^2)
\end{aligned}\]

Considering that $x_{0i}$ and $y_{0i}$ are fixed in the problem (given from a catalog), as well as $x_0$, $y_0$ and the bias $(\varepsilon_x,\varepsilon_y)$, $\textsf{Var}(\nu_i)$ is quadratically derived from Eq.\,\ref{eq:ang_meas}:

\begin{equation}\label{eq:OTopt}
\forall i \in\{1..n^\star\}, \textsf{Var}(\nu_i)=2 \sigma_{in}^4
\end{equation}
Thus, the expression of the rank $n^\star$ matrix $W=(2\sigma_{in}^4)^{-1} \, I_{n^\star}$.

Although easy to compute, $\textsf{Cov}(E)=(C^TWC)^{-1}$ is not analytically simple in the general case. However, in the special case when the searched location of B (foreground body's photocenter) is at the barycenter of a homogeneous cloud of images of background stars, the expression becomes simple. Noting $d$ the standard deviation for both coordinates $x_{0i}$ and $y_{0i}$, and expecting a diagonal matrix $\textsf{Cov}(E)=(\sigma_{\textsf{out}}^2)\,I_2$, we can show after some development that, in this case:

\begin{equation}\label{eq:OTsigma}
\sigma_{\textsf{out}} \simeq \frac{\sigma_{\textsf{in}}}{\sqrt{n^\star}}\left(\frac{\sigma_{\textsf{in}}}{2d}\right).
\end{equation}

This result illustrates how efficient is the presented enhancement of the \mcc. The first improvement factor $1/\sqrt{n^\star}$ is expected as the result of multiple independent raw measurements. The second factor $\sigma_{\textsf{in}}/(2d)$ expresses that the \mcc\ improves the accuracy as a function of the size $2d$ of the cloud of stars in the image compared to the raw accuracy. A numerical application with $\sigma_{\textsf{in}}=10''$, while classical COTS imagers\footnote{e.g. GOM\textsc{space} C1U $19''$ in 35\,mm f/1.9 or $9.5''$ in 70\,mm f/1.9, Berlin Space Technologies IMS-100 $18''$ in 25\,mm, Southern Space Ltd. Gecko $16''$ in 70.5\,mm} have a pixel's angular size ranging from $10''$ to $20''$, with $n^\star=5$ and $2d=1000''$ (100 pixels) yields $\sigma_{\textsf{out}}=0.022''$. The sensitivity of this approximation is still to be studied in the general case: B not at the barycenter, cloud of star images not homogeneous, strong optical aberrations or bias in identifying a reference frame. Nevertheless, we consider in the following, as a conservative approach, that the accuracy of the \mcc\ is better than 0.2$''$. Hence, the use of such an enhanced \mcc\ is in the baseline of our orbit determination solution.

\section{Asynchronous Triangulation}\label{at}

Our orbit determination  is an estimator of the state vector of the CubeSat from the successive directions of various celestial bodies. It performs a progressive triangulation based on non-simultaneous measurements. Any single triangulation would use a minimum set of directions and would yield a poor accuracy. Instead, the observed directions directly feed a Kalman filter that actually performs a triangulation. The geometric relations at stake are strongly non linear, requiring a \emph{sequential} and \emph{unscented} Kalman filter.

\subsection{Observational Model}\label{obsmodel}

\begin{figure}[t]\centering
   \includegraphics[width=0.7\linewidth]{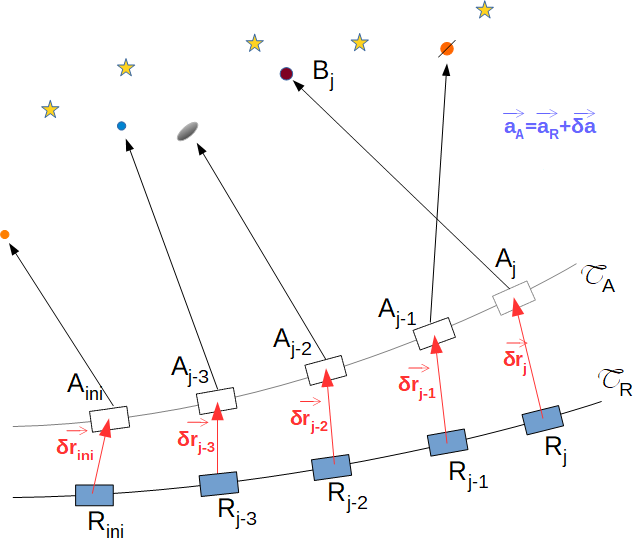}
   \caption{\label{fig:ATnota} Notations for the asynchronous triangulation: ``B'' for celestial bodies, ``R'' for locations on the reference trajectory $T_R$, ``A'' on the actual trajectory $T_A$.}
\end{figure}

We consider that a selection of foreground celestial bodies is available at any time. The availability of the trajectory $T_R$, pre-computed and stored on board, allows to search for a body B in a region of the sky. Then the \mcc\ provides the actual direction of B, with its covariance, as measured by the Object Tracker from an actual location A of the CubeSat on its trajectory $T_A$. Each location A$_j$ corresponds to a reference location R$_j$ on $T_R$ and the estimator will reconstruct the shift of the 9-element state vector $(\vec{\delta r},\vec{\delta v},\vec{\delta a})$ from $T_R$ to $T_A$, as summarized in figure \ref{fig:ATnota}.

An elementary observation is considered to be the shift in elevation $d\varphi$ and in azimuth $d\lambda$ of B between positions on $T_A$ and $T_R$. They are directly taken from the orientation requested to the optical sensor and provided by it after \mcc. Then, at any given measurement $k$:
\begin{equation}\label{eq:ukf_h1}
\begin{pmatrix} d\lambda_k \\ d\varphi_k \end{pmatrix}
= \begin{pmatrix} \lambda_{\textsf{MCC}} - \lambda_{\textsf{requested}} \\
\varphi_{\textsf{MCC}} - \varphi_{\textsf{requested}} \end{pmatrix}
\end{equation}

The covariance of the measurement (Eq.\ref{eq:OTsol}) is taken as the measurement process noise. Both elements define our raw measurement for the non linear Kalman filter:
\begin{equation}\label{eq:ukf_h2}
\begin{aligned}
z_k = ~ & h(x_k,t_k)+v_k, & v_k &\sim\mathcal{N}(0,R_k) \\
\text{with~} & & \\
 & z_k = \begin{pmatrix} d\lambda_k \\ d\varphi_k \end{pmatrix},
 & R_k &= \textsf{Cov}\begin{pmatrix} \lambda_{\textsf{MCC}} \\ \varphi_{\textsf{MCC}} \end{pmatrix}, \\
\end{aligned}
\end{equation}
where the \emph{analytical model} $h(x_k,t_k)$ behind these observations is a direct computation in spherical coordinates of $\overrightarrow{R_kB_k}$ and $\overrightarrow{A_kB_k}$:
\begin{equation}\label{eq:ukf_h3}
h(x_k,t_k)
= \begin{pmatrix} \lambda_{A_kB_k} - \lambda_{R_kB_k} \\
\varphi_{A_kB_k} - \varphi_{R_kB_k} \end{pmatrix}
\end{equation}
with $A_k = R_k+\overrightarrow{\delta r_k}$.

Then, $\overrightarrow{A_kB_k}$ is given by the rectangular coordinates of $R_k$ and~$B_k$:
\[\left\lbrace
\begin{aligned}
\varrho_{R_kB_k} &= \textsf{dist}(R_k,B_k) \\
\varphi_{R_kB_k} &= \arcsin\left(\frac{z_{B_k}-z_{R_k}}{\varrho_{R_kB_k}}\right) \\
\lambda_{R_kB_k} &= \textsf{sign}(\lambda_{R_kB_k}). \\
 &~~ \arccos\left(\frac{x_{B_k}-x_{R_k}}{\varrho_{R_kB_k}\cos\varphi_{R_kB_k}}\right),
\end{aligned}\right.
\]
where $\textsf{sign}(\lambda_{R_kB_k})=\textsf{sign}(y_{B_k}-y_{R_k})$.

And $\overrightarrow{A_kB_k}$ is given in a similar way:
\[
\left\lbrace
\begin{aligned}
\varrho_{A_kB_k} &= \textsf{dist}(R_k+\overrightarrow{\delta r_k},~B_k) \\
\varphi_{A_kB_k} &= \arcsin\left(\frac{z_{B_k}-z_{R_k}-\left(\overrightarrow{\delta r_k}\right)_z}{\varrho_{A_kB_k}}\right) \\
\lambda_{A_kB_k} &= \textsf{sign}(\lambda_{A_kB_k}). \\
 &~~ \arccos\left(\frac{x_{B_k}-x_{R_k}-\left(\overrightarrow{\delta r_k}\right)_x}{\varrho_{A_kB_k}\cos\varphi_{A_kB_k}}\right),
\end{aligned}\right.
\]
where $\textsf{sign}(\lambda_{A_kB_k})=\textsf{sign}(y_{B_k}-y_{R_k}-(\overrightarrow{\delta r_k})_y)$.

Simulations and tests showed that feeding the filter with direct measurements is better than gathering a batch of measurements to first allow a single triangulation and then feed the filter. In addition, this approach is more responsive as the filter is updated after every single measurement and more flexible as it opens to future new types of measurements.

\subsection{Dynamical Model for the \ukf}\label{ukf_m}
The theory of the Kalman filtering builds on the heritage of the estimators of Markov, Kolmogorov and Wiener. It benefits from a rich literature to address many contexts, for instance, by Simon 2006 \cite{Simon2006}. We take from it the components and notations, commonly adopted in the field, for the implementation of the ``unscented" variant of the filter. Although more demanding in terms of computation resource, this variant is preferred to the ``extended" one because of its robustness with regard to the need for trigonometric functions and complicated acceleration fields (where the partial derivative does not combine properly). In addition, it proves to show a good potential for future improvements.

In the ``unscented" variant of the Kalman filter, the dynamical and observational models, $f(x_k,u_k,t_k)$ and $h(x_k,t_k)$, are not necessarily linear:
\[
\begin{aligned}
x_{k+1} &= f(x_k,u_k,t_k)+w_k, & w_k &\sim\mathcal{N}(0,Q_k) \\
z_k &= h(x_k,t_k)+v_k, & v_k &\sim\mathcal{N}(0,R_k) \\
\end{aligned}
\]
where the processes noises $w_k$ and $v_k$ are assumed to be zero-mean Gaussian random variables ($h$ and $R_k$ already defined in section \ref{obsmodel}). Then, we expect the filter to produce an \emph{a posteriori estimate} of the state vector at iteration $k$ with its covariance, respectively noted $\hat x_k^+$ and $P_k^+$.

The state vector $x_k$ is taken as a dimensionless 9-element transform from $(\vec{\delta r_k},\vec{\delta v_k},\vec{\delta a_k})$ expressing the shift from $T_R$ to $T_A$. Indeed a scaling matrix $S$ was deemed necessary to avoid numeric degeneracy. It is built as follows:
\begin{equation}\label{eq:S_matrix}
\begin{matrix}
{S = \begin{bmatrix}
f_x\,I_3 & 0_3 & 0_3 \\
0_3 & f_v\,I_3 & 0_3 \\
0_3 & 0_3 & f_a\,I_3 \end{bmatrix}_{(9\times9)},} \\ {
~\textsf{with}\left\lbrace\begin{aligned}
f_t &\textsf{ and } f_x \textsf{ arbitrary set} \\
f_v &= f_x / f_t \\
f_a &= f_v / f_t
\end{aligned}
\right.}
\end{matrix}
\end{equation}
where $I_3$ and $0_3$ are the $3\times3$ Identity and Null matrices and $f_t$, $f_x$, $f_v$, $f_a$ respectively express a scaling factor for durations, distances, velocities and accelerations, respectively. Hence, the following relations apply to re-scale the state vector and its covariance from their dimensionless expression:
\begin{equation}\label{eq:scaling}
\begin{aligned}
\begin{pmatrix}
\overrightarrow{\delta r_k} \\ \overrightarrow{\delta v_k} \\ \overrightarrow{\delta a_k} \end{pmatrix} &= S \, \hat x_k^+ \\
P_{k,re-scaled}^+ &= S \, P_k^+ \, S
\end{aligned}
\end{equation}

\noindent Note: The scaling matrix $S$ has also to be used for the state vector in the observational model in Eq. \ref{eq:ukf_h3}: the function $h$ accounts for the link from the observables to $(\vec{\delta r_k},\vec{\delta v_k},\vec{\delta a_k}) = S \, x_k$, i.e. \emph{after} re-scaling.

The dynamical model (also called ``process'' model in the main theory of Kalman filtering) is taken as simple as possible, to show results even before improvements on the models. Indeed, it can be much more sophisticated, without changing the presented algorithm, provided that the entire dynamical model is expressed by the function $f(x_k,u_k,t_k)$: for instance, it could take solar radiation pressure into account, or trajectory correction maneuvers with the ``command vector" $u_k$, it could also be a complex trajectory propagator like a Runge-Kutta at high order. Here, the propagation is at order 0 (constant velocity and acceleration), the acceleration being estimated from a non-linear model of the gravitational environment. Also, there is no use of the command vector $u_k$. The estimates from an iteration to the next are given by
\begin{equation}\label{eq:ukf_f}
\begin{aligned}
\hat{x}_k &= f(\hat{x}_{k-1},u_k,t_k) \\
&= \begin{bmatrix}
I_3 & (dt_k/f_t)\,I_3 & 0_3 \\
0_3 & I_3 & (dt_k/f_t)\,I_3 \\
0_3 & 0_3 & I_3 \end{bmatrix} \, \hat{x}_{k-1}
\end{aligned}
\end{equation}

The model is not linear because the local acceleration model $\overrightarrow{a_{loc}}(R_k+\overrightarrow{\delta r_{k-1}})$ is not uniform. In the test case presented in section \ref{case}, the ``on-board'' gravitational model includes the distances to the Sun, the Earth and Mars. It is also less sophisticated than the environment used to produce $T_A$, which considers Jupiter in addition.

The covariance of the process noise $Q_k$ expresses the cumulative uncertainty from iterations $k-1$ to $k$ resulting from the imperfect knowledge of the acceleration. Here, we consider a constant and uniform uncertainty in the acceleration, noted $\sigma_a$. Hence, the dimensionless covariance writes:
\begin{equation}\label{eq:ukf_q}
\begin{aligned}
Q_k &= S^{-1} \, V_k \, S^{-1} \text{, with} \\
V_k &= \begin{bmatrix}
(\sigma_a dt_k^2/2)^2.I_3 & 0_3 & 0_3 \\
0_3 & (\sigma_a dt_k)^2.I_3 & 0_3 \\
0_3 & 0_3 & \sigma_a^2.I_3
\end{bmatrix}.
\end{aligned}
\end{equation}

\subsection{\ukf\ implementation}\label{ukf}

The \ukf\ works with the use of a set of so-called ``sigma-points" that build a mesh around the 9-dimension state vector. We again refer to Simon \cite{Simon2006} for theory, notations and steps in this section. As a summary, a mesh is built from the estimate $\hat x_{k-1}^+$ and its covariance $P_{k-1}^+$ at iteration $k-1$. It is transformed through the dynamical and the observational models. Its resulting new meshes serve to build the a posteriori estimate $\hat x_k^+$ and its covariance $P_k^+$. In our case, the state vector is made of $n=9$ components, hence the number of $2n=18$ needed sigma-points.

The initialization of the filter is taken from $T_R$, i.e. $\hat x_0^+ = 0_{9\times1}$ and with the initial values of the covariance $P_0^+$
\[\begin{array}{l}
P_0^+ = \\
 \begin{bmatrix} (\sigma_x/f_x)^2.I_3 & 0_3 & 0_3 \\ 0_3 & (\sigma_v/f_v)^2.I_3 & 0_3\\ 0_3 & 0_3 & (\sigma_a/f_a)^2.I_3 \end{bmatrix}, \\
\end{array}\]
where $f_x$, $f_v$, $f_a$, $\sigma_a$ were already defined and $\sigma_x$, $\sigma_v$ express an expected maximum shift in distance and velocity of $T_A$ from $T_R$.

Then the detailed implementation consists of nine steps:
\begin{enumerate}
\item A matrix square root of $(n.P_{k-1}^+)$ is computed to dispatch the sigma-points around $\hat x_{k-1}^+$ within the $n$-dimension space. Such a $\Sigma$ matrix may not be unique, provided that $\Sigma^T\,\Sigma=(n.P)$. It may contain complex numbers whose only real parts are kept.
\[
\Sigma_{k-1}^+ = \sqrt{n.P_{k-1}^+}
\]

\item $2n$ sigma-points $\hat{x}_{k-1}^{(i)}$ are built from the last available estimate $\hat{x}_{k-1}^+$:
\[
\begin{array}{l}
i=1..2n, ~ \hat{x}_{k-1}^{(i)} = \hat{x}_{k-1}^+ + \tilde{x}^{(i)}, \\ \\
i=1..n, ~ \left\lbrace \begin{aligned}
\tilde{x}^{(i)} &= \left(\Sigma_{k-1}^+\right)_i^T\\
\tilde{x}^{(n+i)} &= -\left(\Sigma_{k-1}^+\right)_i^T ~,
\end{aligned} \right.
\end{array}
\]
where $\left(\Sigma_{k-1}^+\right)_i$ defines the line $i$ of $\Sigma_{k-1}^+$.

\item The sigma-points are propagated through the dynamical model. Then, an \emph{a priori} new estimate is built at the barycenter of their transform:
\[
\hat{x}_k^{(i)} = f(\hat{x}_{k-1}^{(i)}, u_k, t_k) ~ \textsf{ (ref. Eq. \ref{eq:ukf_f})},
\]
then
\[
\hat{x}_k^- = \frac{1}{2n} \sum_{i=1}^{2n} \hat{x}_k^{(i)}.
\]

\item A new \emph{a priori} covariance is approximated in a similar way:
\[\begin{array}{l}
P_k^- = \\
 \frac{1}{2n} \sum_{i=1}^{2n}
\left(\hat{x}_k^{(i)}-\hat{x}_k^-\right)
\left(\hat{x}_k^{(i)}-\hat{x}_k^-\right)^T
\\ ~+~Q_{k-1}.
\end{array}\]

\item A new mesh of sigma-points replaces the previous mesh $\hat{x}_k^{(i)}$ by taking advantage of the latest a priori estimate and covariance:
\[
\begin{array}{l}
i=1..2n, ~ \hat{x}_k^{(i)} = \hat{x}_k^- + \tilde{x}^{(i)}, \\
\Sigma_k^- = \sqrt{n.P_k^-} \\
i=1..n, ~ \left\lbrace \begin{aligned}
\tilde{x}^{(i)} &= \left(\Sigma_k^-\right)_i^T\\
\tilde{x}^{(n+i)} &= -\left(\Sigma_k^-\right)_i^T ~.
\end{aligned} \right.
\end{array}
\]

\item The mesh is transformed in observables $\hat{z}_k^{(i)}$ and the barycenter of this transform yields a prediction of measurement $\hat{z}_k$:
\[
\hat{z}_k^{(i)} = h(\hat{x}_k^{(i)}, t_k) ~ \textsf{ (ref. Eq. \ref{eq:ukf_h3})},
\]
then
\[
\hat{z}_k = \frac{1}{2n} \sum_{i=1}^{2n} \hat{z}_k^{(i)}.
\]

\item In parallel, $P_k^-$ is transformed into an observable covariance $P_z$, after adding the measurement process noise $R_k$ (from Eq. \ref{eq:ukf_h2}), as a $2\times2$ matrix:
\[\begin{array}{l}
P_z = \\
 \frac{1}{2n}.\sum_{i=1}^{2n} 
\left(\hat{z}_k^{(i)}-\hat{z}_k\right)
\left(\hat{z}_k^{(i)}-\hat{z}_k\right)^T
+R_k.
\end{array}\]

\item A cross-covariance is defined as a $9\times2$ matrix, between the estimate and its transform into an expected observable:
\[\begin{array}{l}
P_{xz} = \\
 \frac{1}{2n}.\sum_{i=1}^{2n} 
\left(\hat{x}_k^{(i)}-\hat{x}_k^-\right)
\left(\hat{z}_k^{(i)}-\hat{z}_k\right)^T .
\end{array}\]

\item Eventually, the update of the estimate with its covariance is possible, setting a Kalman gain as a $9\times2$ matrix noted $K_k$:
\[
\begin{aligned}
K_k &= P_{xz} \, P_z^{-1}, \\
\hat x_k^+ &= \hat x_k^- + K_k (z_k - \hat z_k), \\
P_k^+ &= P_k^- - K_k \, P_z \, K_k^T .
\end{aligned}
\]
\end{enumerate}

The process involves only one $2\times2$ matrix inversion $P_z^{-1}$. However, two $9\times9$ matrices square root $\Sigma$ have also to be computed. A possible algorithm is suggested by Sadeghi (\cite{Sadeghi2018}, 2018), for instance, and some well-validated librairies may be considered like the \gls{GSL} for C/C++.

\section{Results}\label{results}

We present the performance for a specific case study of a cruise from Earth to Mars. This case study does not allow to conclude for all contexts of use but it illustrates the efficiency of the algorithm and provides good insights for other contexts. We also present an initial assessment of the CPU considering CubeSat hardware as a target.

\subsection{Study case}\label{case}
\begin{figure}[t]\centering
 \includegraphics[width=0.45\linewidth]{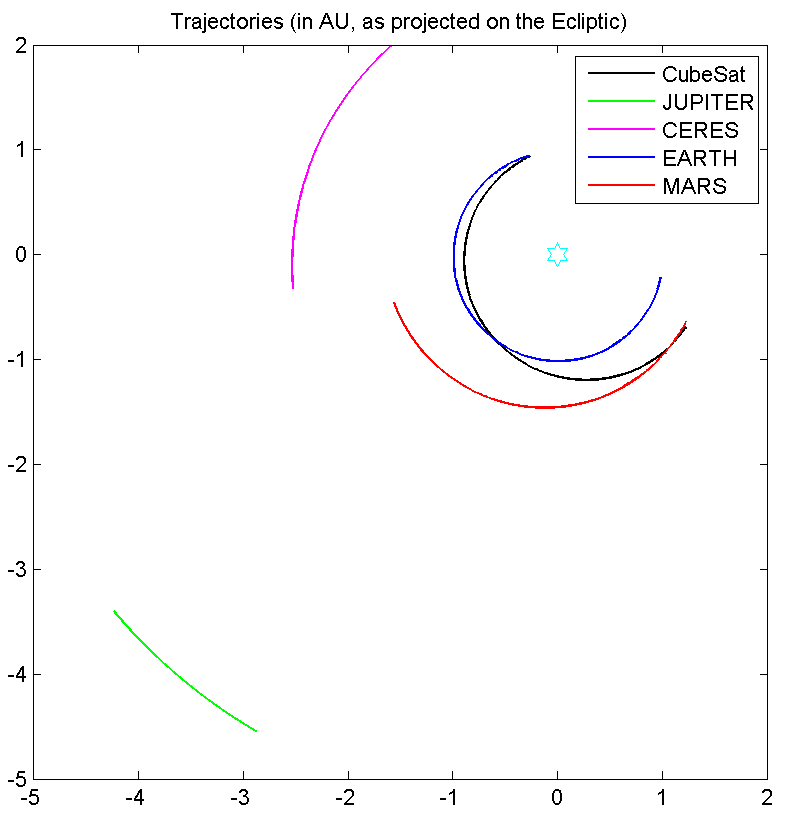}
 \includegraphics[width=0.5\linewidth]{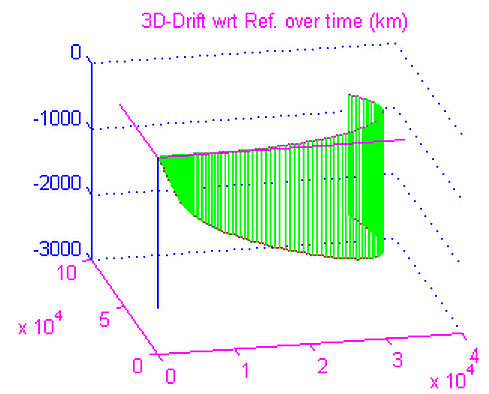}
   \caption{\label{fig:E2M}CubeSat and celestial bodies trajectories in 2018 (left, unit AU); Shift of $T_A$ from $T_R$ (right, unit km).}
\end{figure}

A cruise flight from Earth to Mars is assumed: $T_R$ is propagated from the injection into interplanetary orbit up to a flyby of Mars, simulating a host mission launched in 2018; $T_A$ is then propagated with $T_R$'s initial conditions after adding a small retrograde $\Delta V$ of 1\,m/s, simulating the effect of jettisoning a CubeSat. The drift of $T_A$ from $T_R$ expands up to $\sim$40\,000\,km over the journey (Fig.\,\ref{fig:E2M}). It allows to assess the accuracy of the presented Orbit Determination (OD) in a vast domain of use.

The OD process is ``run'', i.e. re-started, every 10 days during the journey, allowing an analysis of its behavior all over the period. When the process starts, it first initializes the filter (section \ref{ukf}), then performs 600 filter's updates (``iterations''), taking 1 day and 3 hours of flight. We also tested ``runs'' that include 8000 iterations that correspond to 15 days of flight.

Each iteration simulates a new optical observation processed with the \mcc\ in the following loop sequence: 3 images to the Earth in 3 minutes, a 5-mn slew, 3 images to Mars, a 5-mn slew, 3 images to Ceres, a 5-mn slew. As it was shown in section \ref{sensor}, various solutions for celestial bodies are available and we consider here the Earth and Mars as being unresolved, despite they may not be. Other bodies may be used alternatively and they do not impact the interpretation of the filter's behavior. However, it provides interesting lessons about the filter at the beginning and at the end of the journey that can be discussed further. Then, a simulated observation consists of the computation of the actual direction of the celestial body as seen from $T_A$ with an additive Gaussian noise of 0.2$''$ standard deviation, as it would result from the \mcc. The expected directions as seen from $T_R$ are computed from assumed well-known ephemerides of the bodies that are considered available on board. The observations and the transform of sigma-points into observables are performed according to Eq. \ref{eq:ukf_h2} and \ref{eq:ukf_h3}. The physical evolution of the mesh of sigma-points is considered by the modeling of the local gravitational contributions of the Sun, the Earth and Mars all over the scenario.

\subsection{Orbit Determination Accuracy}\label{od}

\begin{table}[t]
  \begin{center}
    \caption{\label{tbl:results} Main results for runs of 600 and 8000 iterations, re-started every 10 days over a scenario of a 230-day journey from Earth to Mars}
    \begin{tabular}[width=\linewidth]{|l|c|c|}
      \hline
      \textbf{Runs} & \textbf{600-it.} & \textbf{8000-it.}\\
      \hline
      duration (days) & 1.125 & 15\\
      \hline
      convergence (days) & 0.9 & $\sim1$ \\
      \hline
      3-$\sigma$ enveloppe (km) & 30...150 & 30...150 \\
      \hline
      Best epoch & day 180 & day 180 \\
      \hline
    \end{tabular}
  \end{center}
\end{table}

\begin{figure}[t]
\includegraphics[width=\linewidth]{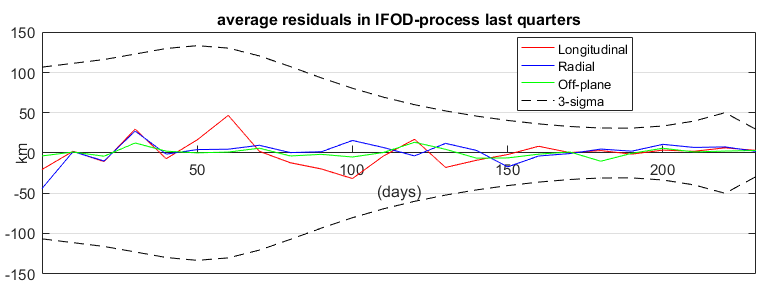}
\caption{\label{fig:UKF_1}OD accuracy. Residuals (in km) at the end of each run (1 run takes 600 iterations and lasts $\sim1\,$day), a new run is re-started every 10 days. The dashed line shows an envelope of the 3-$\sigma$ error bars. $\sigma_{\textsf{MCC}}=0.2$".}
\end{figure}

\begin{figure*}[t]
\includegraphics[width=\linewidth]{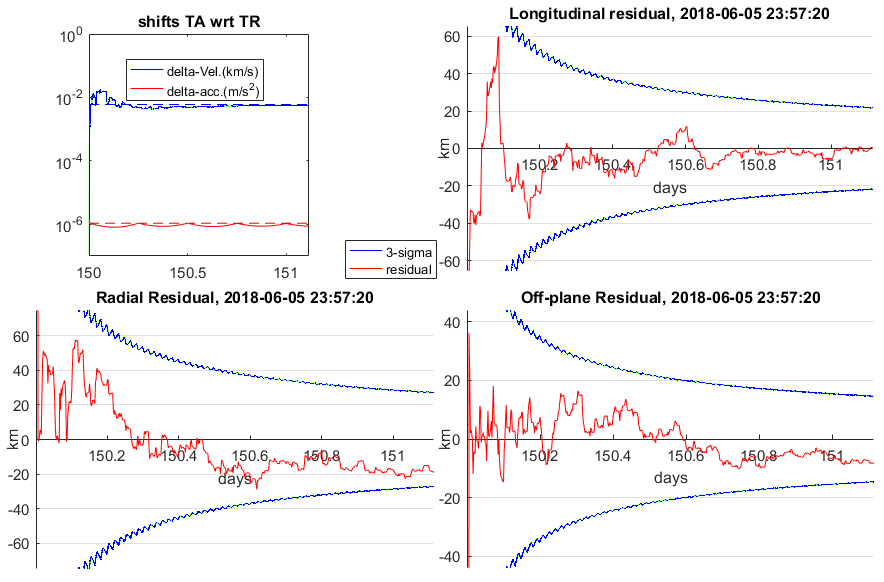} \caption{\label{fig:UKF_2}Details at day 150 of the \ukf\ behavior. Convergence of the filter for velocities in km/s and accelerations in km/s$^2$ (top left), residuals in longitudinal (top right), radial (bottom left), off-plane (bottom right) directions. $\sigma_{\textsf{MCC}}=0.2$".}
\end{figure*}

A synthesis of the results is given in Fig.\,\ref{fig:UKF_1} and Table\,\ref{tbl:results}. The detailed behavior of a particular run is presented in Fig.\,\ref{fig:UKF_2} for day 150 within the scenario. All results of the \ukf\ are obtained in dimensionless values and re-scaled. Then, the standard deviations are taken from the diagonal terms of the covariances, the distance terms being first transformed by a change of frame to have them expressed in the heliocentric TNW frame (tangent, normal, momentum, or longitudinal, radial and off-plane) attached to the trajectory.

In Fig.\,\ref{fig:UKF_1}, the dashed line envelope is defined as the average of the 3-$\sigma$ resulting from the filter during the last quarter of iterations, i.e., iterations 450 to 600 of the same run. The colored plain lines show the OD residuals at iteration 600 in longitudinal, radial and off-plane directions. The OD residual is always included in the envelope; this shows the filter's consistency. The 3-$\sigma$ accuracy is always better than 150\,km and even below 30\,km in the third part of the scenario. The variations illustrate the changing geometry of the directions of the celestial bodies over the scenario and its effect on the filter. The thorough study of optimized geometrical configurations could allow, for instance, in mission preparation, to decide the best period in the cruise to perform the OD or the best set of foreground celestial bodies at a given time of the journey.

The details for day 150 in Fig. \ref{fig:UKF_2} helps understand how the filter works. The 9-element state vector converges in less than 0.2 day for velocities and accelerations (top left, the dashed and plain lines show the expected and reconstructed values of $T_A$ wrt $T_R$). The location residuals (3-$\sigma$ green envelopes) converge in less than 1 day. A longer duration of the filter would not significantly improve the convergence further. At that specific epoch (day 150), the longitudinal residuals converge faster than the radial ones. But this is not always the case, it depends on the changing geometry over the scenario. However, the off-plane residuals are always better.

One can note that the 3-$\sigma$ envelopes show a see-saw shape. It results from the observation strategy with three measurements in a row (1 per minute) of a celestial body before slewing to the next. It illustrates that, when the geometry has not changed from the first to the second and third observations, the quality of the estimate relies mostly on the dynamical model and degrades over time due to the process noise. Such redundant observations should be avoided, unless cases where an image shows a bad quality, and could save observation time by a factor of $\sim25\%$ (6\,min instead of 8\,min per body, including slew). Also if multiple imagers are available or if the \ADCS\ allows quicker slews, slew time can be saved. However, the filter \emph{needs} the geometry of the celestial bodies to evolve from one iteration to the next because uncertainties accumulate over time. It even provides a criterion for an observation strategy: two pictures of the same celestial body are useful at two successive epochs if its direction in the meantime has changed by more than the accuracy of the \mcc.

\begin{figure*}[t]
\includegraphics[width=\linewidth]{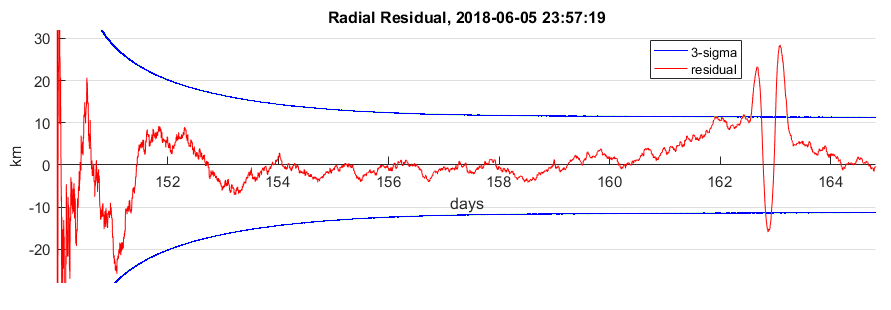}
\caption{\label{fig:UKF_3}Detail at day 150 over a run of 8000 iterations ($\sim15$\,days). $\sigma_{\textsf{MCC}}=0.2$".}
\end{figure*}

In terms of ultimate performance of the filter, some divergence is reported in long runs that include 8000 iterations. It may be a temporary divergence that looks like the filter reaches a singularity and re-starts nominally afterwards (see day 163 in Fig. \ref{fig:UKF_3}). In this case, the divergence goes beyond the 3-$\sigma$ envelope, remaining below 30\,km, but it may also diverge beyond the 3-$\sigma$ envelope (a 60\,km residual is reported while 3-$\sigma\sim15$\,km), then converging again but slower. This behavior is not well-explained and may be related to an effect of filter ``aging". At least it does not come from the geometry: indeed, the run with 8000 iterations takes 15 days (until day 165), overlapping so with the simulation of the next run that starts 5 days earlier (day 160), and the latter does not show any divergence at day 163.

On the opposite, a performance better than the kilometer (residual and 3-$\sigma$) was also observed at the flyby of Mars. Although the observation conditions are not operational (the planet would be resolved on the images), this result illustrates the excellent behavior of the \ukf\ when linear approximations are not possible. Indeed, the flyby at Mars was noticed incompatible with a past linear variant of the filter. The reason was the inapplicable Taylor series of the sine and cosine in the observation model and the approximation of constant local acceleration in the dynamical model. In the new ``unscented" variant, the same situation provides excellent results, getting rid of any requirements on the geometry and the acceleration field for applicability of the OD.

\subsection{Compatibility with on-board CPU}\label{cpu}

The overall CPU for a single iteration of the presented algorithm is estimated lower than $\sim1$\,s on typical CubeSat hardware. It does not include the image processing presented for the \mcc\ that comes in addition (if performed by the same CPU) or in parallel (if performed by the sensor or a separate microcontroller).

The presented results were obtained with a high-level open-source language adapted to prototyping and analysis, making sure that no proprietary coding is required (GNU OCTAVE, however intensive computation was run with MATLAB by Mathworks$^\textsf{{\textregistered}}$ for higher simulation speed). Such language is likely not transposable to CubeSat hardware. We assume here that the on-board code is written in C/C++ with the use of the GNU Scientific Library (GSL).

Before any consideration about the coding, we present an estimate of the CPU based on a careful breakdown in elementary operations. The CPU for these operations was then measured with an Arduino-like hardware to get an estimate of the CPU for the whole algorithm. The values are gathered in Table \ref{tbl:dureesCPU}, considering the following characteristics:
\begin{itemize}
\item the hardware target is a STM32F407VGT6 microcontroller, by ST Micro based on ARM$^\textsf{{\textregistered}}$ Cortex$^\textsf{{\textregistered}}$-M4 32-bit RISC core operating at a frequency of up to 168 MHz
\item elementary operations are identified in 3 main categories: ``bo" for basic operations (e.g. additions), ``co" for single complex operations (e.g. multiplications or divisions of real numbers), ``to" for multiple complex operations, typically when Taylor series developments are assumed (e.g. sine, cosine)
\item a highly consuming operation is identified, a $9\times9$ matrix square root, that could not be tested with the STM32, it is noted ``msr" and is assumed to be of the same complexity as a rank 9 matrix inversion whose CPU was measured instead.
\item the duration for each category was measured separately, from the simplest to the most complicated: for each category, a large number of elementary operations were run, a fit was performed to find the CPU cost of a category while removing the costs for already determined operations and any non-recurring new cost.
\end{itemize}

\begin{table}[t]
  \begin{center}
    \caption{\label{tbl:dureesCPU} CPU estimate for 1 \ukf\ iteration. Breakdown in operation categories of increasing complexity: bo, co, to, msr.}
    \begin{tabular}[width=\linewidth]{|l|c|c|c|c|}
      \hline
      \textbf{Operations} & \textbf{bo} & \textbf{co} & \textbf{to} & \textbf{msr}\\
      \hline
      duration (ms) & 0.012 & 0.079 & 0.15 & 56.4 \\
      \hline
      number & 5~968 & 5~688 & 127 & 2 \\
      \hline
      Total (ms) & 69.7 & 449 & 19.2 & 113 \\
      \hline
      \hline
    \end{tabular}
    \begin{tabular}[width=\linewidth]{|l|c|}
      \textbf{Full iteration} & \textbf{$\sim650$\,ms}\\
      \hline
    \end{tabular}
  \end{center}
\end{table}

These numbers must be considered with caution since some considerations were not detailed here as we only expect a magnitude of the consumed CPU: hence, double or quadruple precision, access optimization to RAM and mass storage, for instance, are not considered. In return, the STM32 is considered well below the capacities of nowadays CubeSat boards, processing often at 400 or 800\,MHz and arranged with hundreds of MB in RAM and mass storage. Hence, a CPU budget of 1\,s is considered reasonable with a comfortable margin. It even opens to further improvements in the algorithm.

\section{Discussion and prospects}\label{next}

\subsection{\mcc\ at TRL\,2}
The major source of error is the optical sensor and the \mcc\ algorithm is the key to reach an optical accuracy better than 0.2\,". As an assessment, the \mcc\ must be considered as a technological brick of the solution at \TRL\,2 based on the present paper and a few concerns are still to be studied.

Centroiding can offer a precision up to one half to one tenth of a pixel, whose angular size is typically at 10\,". Other errors come into play: a small aperture is desirable to get a larger PSF that, however, remains much smaller than a pixel (0.2\," in visible light with a 1-cm aperture) making the centroiding poorly efficient with faint objects at the limit of sensitivity.

Optical aberrations and change of frame between celestial directions and coordinates in the field of view introduce alignment errors. These errors can be partially corrected based on calibration and re-centroiding within the \mcc\ itself, but their effect on the performance of the \mcc\ is still to be assessed.

A catalog of single bright stars is required in mission preparation. Their parallaxes must be included as we deal with interplanetary journeys. Stars weaker than the sensitivity or too close from one another must be removed to ensure that only one star can be identified per region of interest. 

\subsection{Observational strategy}
The performance of the \ukf\ is essentially driven by the availability of celestial bodies, and by an observational strategy to decide the sequence of observations.

The careful selection of such bodies in mission preparation is possible: criteria like the visibility and the magnitude of planets' moons, asteroids or even the host mission or other visible spacecrafts, their elongation to the Sun, the accuracy of their ephemerides, the availability of cataloged stars in background can be combined and can evolve over the expected journey.

We mentioned a criterion to decide for an additional measurement of a body when it moved by at least $\sigma_\textsf{MCC}$ which can be anticipated in the on-board observation schedule. However its contribution to the improvement of the estimate is not known before the direction is actually measured, meaning to point at it which has a non-negligible ADCS cost. Considering that the geometry of stars and bodies evolve slowly, a machine learning process could likely take advantage of the first tens of measurements to optimize the \ADCS\ cost by allowing the observations that will most decrease the state vector's covariance among all next possible observations.

\subsection{\ukf\ potential}
New kinds of observations could also be introduced in the orbit determination algorithm while keeping its structure.

The tolerance for non-linear models allows multiple observations of different kinds in parallel. For instance, images of resolved objects could be considered, like a close moon or planet with its phase effect or an asteroid in proximity operations. Such observations require a specific algorithm (e.g. Modenini et al. \cite{Modenini2018}). If a direction is output by the algorithm, the rest of the \ukf\ is usable as is. The output may be different from a direction, for instance, with algorithms for horizon-based navigation that derive observations \emph{not} in terms of elevation and azimuth. Then, a dedicated observation model is to be developed and can run beside the model used in section \ref{obsmodel}. The structure of the \ukf\ is easy to adapt for multiple concurrent observation models, by simply switching at every iteration to the wanted model for $z_k$, $h(x_k,t_k)$ and the evaluation of sigma-points ($\hat{z}_k^{(i)}$, $P_z$, $P_{xz}$). The increase in computing resources will be due to storage needs for a larger code and processing needs for the inversion of the largest matrix $P_z$ of a particular  model.

Observations not sorted in chronological order can still feed the \ukf, allowing measurements to be received with a delay. This is desirable with, for instance, radio-science tracking (ranging and Doppler) from the ground or a mothercraft in proximity operations. Such measurements can be transmitted afterward, not in real-time, and arrived after younger on-board observations have been already processed by the \ukf. Literature about the theory of Kalman filtering, e.g. Simon \cite{Simon2006}, is available to deal with estimate updates based on past measurements. These additions for proximity operations, as fully orthogonal measurements, would for sure dramatically improve the performances of the on-board orbit determination at a minor cost for ground operations.

\section{Conclusions}

Current developments in autonomous navigation are part of a broad endeavor to use CubeSats in deep space. We contribute to this effort with a plausible solution adapted to this scale.

We present an affordable solution for orbit determination that can reach up to a 30\,km accuracy (3-$\sigma$) in a cruise context, at a reasonable system cost in terms of \ADCS, observation duty-cycle and CPU. This solution relies on a mission preparation to assess the possible observations and the best period of the journey for the orbit determination. It is based on various assumptions: observations of foreground objects like optically unresolved planets or their major moons, possibly bright asteroid when not enough planets or moons can be observed, optical sensor with a sensitivity down to $m_V=7$ at least. With these assumptions, the solution includes an image processing called ``enhanced \mcc'' (Multiple Cross-Correlation) that is specified here, bringing it at \TRL\,2. The \mcc\ allows, in the presented conditions, to measure a direction of a celestial body with an accuracy better than $0.2''$ with typical CubeSat hardware.

We show that the solution can be further adapted to contexts like proximity operations that do not tolerate linear approximations. Moreover, the algorithm is ready for data fusion of a stream of multiple kinds of observations, even if not provided in their chronological order. As well, further improvements are identified to save \ADCS\ resources and speed up the process by a machine learning that would train during the first iterations of the filter.

\section*{Acknowledgment}

An extensive development of the work summarized here is available in French in Segret 2019 \cite{Segret2019a} that benefited from the thorough support of Pr. Véronique \textsc{Dehant} at the Royal Observatory of Belgium (ROB); Dr. Mathieu \textsc{Barthélémy} at the Institute for Planetary sciences and Astrophysics (IPAG), Observatory of Grenoble; Pr. Pierre \textsc{Drossart} and M.Sc. Philippe \textsc{Plasson} at LESIA, Observatory of Paris. The science cases were kindly provided by Pr. Daniel \textsc{Hestroffer} of IMCCE, Observatory of Paris, for the case in planetary geodesy and by Dr. Jordan \textsc{Vannitsen} of \textsc{Odysseus Space SA} for the case in space weather. We also thank M.Sc. Rashika Jain at LESIA, Observatory of Paris, for the help in clarifying the paper.

This work was funded by the French Laboratory of Excellence ESEP [grant ANR 2011-LABX-030 in ``Investissements d’Avenir"]; and Université PSL (Paris Sciences et Lettres), France.

\appendix
\section*{Appendix. Simulation data-set}

A data-set is provided with the assumptions and results:
\begin{itemize}
\item For convenience, a subfolder is provided and ready-to-use with a copy of the 600-iteration set and organized to be displayed by French space agency CNES' free software VTS for space data visualization (ref.: \textsf{https://timeloop.fr/}). The VTS configuration file to be open is ``IFOD\_EME.vts".
\item Input files consist in ascii files:
	\begin{itemize}
	\item trajectories $T_R$ (``BIRDY\_TR.xva") and $T_A$ (``BIRDY\_TA\_j-1.xva")
	\item ephemerides of the Earth, Mars, Ceres as extracted from Paris Observatory's IMCCE using INPOP13C planetary theory, and re-formated
	\end{itemize}
\item Output files contain 2 sets, for runs with 600 and 8000 iterations, each set is composed of the following files:
	\begin{itemize}
	\item ``...\_OD\_TRAJ.bin" and ``...\_ELOIDS.cvbin" are binary files that contain the raw results of Orbit Determination at every single iteration, i.e. the state-vector (as a shift from $T_R$) and its covariance respectively,
	\item ascii files ``...\_OD\_TRAJ" and ``...\_ELOIDS", same like above, adapted for direct display with VTS,
	\item ascii file ``...\_LRESIDL", ``...\_TRESIDL" are the residuals in longitudinal and transverse directions,
	\item ascii file ``...\_INOUT3S" is a color-code file for VTS that summarizes the result of the Orbit Determination (grey, green, red or orange),
	\item ascii files ``...\_TGTBODY" is a synthetic file of the directions of the foreground bodies to look at over time, for orientations is VTS display.
	\end{itemize}
\item Dimensionless parameters and accuracy values at start:
\[
\left\lbrace\begin{aligned}
f_t &= 100\,\textsf{s} \\
f_x &= 30\,000\,\textsf{km} \\
f_v &= f_x / f_t \\
f_a &= f_v / f_t 
\end{aligned}\right.
~ \left\lbrace\begin{aligned}
\sigma_x &= 80\,000\,\textsf{km} \\
\sigma_v &= 0.015\,\textsf{km/s} \\
\sigma_a &= 4.10^{-9}\,\textsf{km/s}^2 
\end{aligned}\right.
\]

\end{itemize}

  \bibliography{MyCollection}

\end{document}